\documentclass[twocolumn]{aastex62}
\usepackage{amsmath}
\usepackage{float}

\graphicspath{{./}{Figures/}}

\begin{document}
	
	\title[Puffy accretion disks]{Puffy accretion disks: sub-Eddington, optically thick, and stable}
	
	\correspondingauthor{Debora Lan{\v c}ov\'a}
	\email{debora.lancova@fpf.slu.cz, wlodek@camk.edu.pl,\\maciek.wielgus@gmail.com}
	
	\author{Debora Lan{\v c}ov\'a}
	\affiliation{Research Center for Computational Physics and Data Processing; Research Centre of Theoretical Physics and Astrophysics,\\ Institute of Physics, Silesian University in Opava, 
	Czech Republic}
	\affiliation{Black Hole Initiative at Harvard University, 20 Garden Street, Cambridge, MA 02138, USA}
	
    \author{David Abarca}
	\affiliation{Nicolaus Copernicus Astronomical Centre, Polish Academy of Sciences, Bartycka 18, 00-716 Warsaw, Poland}
        
    \author{W{\l}odek Klu{\'z}niak}
    \affiliation{Nicolaus Copernicus Astronomical Centre, Polish Academy of Sciences, Bartycka 18, 00-716 Warsaw, Poland}
				
	\author{Maciek Wielgus}
	\affiliation{Black Hole Initiative at Harvard University, 20 Garden Street, Cambridge, MA 02138, USA}
		\affiliation{Center for Astrophysics  $|$ Harvard \& Smithsonian, 60 Garden Street, Cambridge, MA 02138, USA}
	\affiliation{Nicolaus Copernicus Astronomical Centre, Polish Academy of Sciences, Bartycka 18, 00-716 Warsaw, Poland}
    
    \author{Aleksander S{\c a}dowski}
    \affiliation{Akuna Capital, 333 South Wabash Av., Chicago, IL 60604, USA}
	
	\author{Ramesh Narayan}
  	\affiliation{Center for Astrophysics  $|$ Harvard \& Smithsonian, 60 Garden Street, Cambridge, MA 02138, USA}
	\affiliation{Black Hole Initiative at Harvard University, 20 Garden Street, Cambridge, MA 02138, USA}
	
	\author{Jan Schee}
    \affiliation{Research Center for Computational Physics and Data Processing; Research Centre of Theoretical Physics and Astrophysics,\\ Institute of Physics, Silesian University in Opava, 
	Czech Republic}
        
    \author{Gabriel T\"{o}r\"{o}k}
    \affiliation{Research Center for Computational Physics and Data Processing; Research Centre of Theoretical Physics and Astrophysics,\\ Institute of Physics, Silesian University in Opava, 
	Czech Republic}
  	
	\author{Marek Abramowicz}
	\affiliation{Research Center for Computational Physics and Data Processing; Research Centre of Theoretical Physics and Astrophysics,\\ Institute of Physics, Silesian University in Opava, 
	Czech Republic}
	\affiliation{Nicolaus Copernicus Astronomical Centre, Polish Academy of Sciences, Bartycka 18, 00-716 Warsaw, Poland}
	\affiliation{Department of Physics, G{\"o}teborg University, Sweden }
	\affiliation{Black Hole Initiative at Harvard University, 20 Garden Street, Cambridge, MA 02138, USA}

\begin{abstract}
We report on a new class of solutions of black hole accretion disks that we have found through three-dimensional, global, radiative magnetohydrodynamic simulations in general relativity.
 It combines features of the canonical thin, slim and thick disk models but differs in crucial respects from each of them. We expect these new solutions to provide a more realistic description of black hole disks than the slim disk model. We are presenting a disk solution for a non-spinning black hole at a sub-Eddington mass accretion rate, $\dot M=0.6\,\dot M_{\rm Edd}$. By the density scale-height measure the disk appears to be thin, having a high density core near the equatorial plane of height $h_{\rho} \sim 0.1 \,r$, but most of the inflow occurs through a  highly advective, turbulent, optically thick, Keplerian region that sandwiches the core and has a substantial geometrical thickness comparable to the radius, $H \sim r$. The accreting fluid is supported above the midplane in large part by the magnetic field, with the gas and radiation to magnetic pressure ratio $\beta \sim 1$, this makes the disk thermally stable, even though the radiation pressure strongly dominates over gas pressure. A significant part of the radiation emerging from the disk is captured by the black hole, so the disk is less luminous than a thin disk would be at the same accretion rate.

~~~~~~~~~~~~~~~~~~~~~~~~~~~~~~~~~~~~~~~~~~~~~~~~~~~~~~~~~~~~~~~~~~~~~~~~
	
\end{abstract}
	
	
        \section{Introduction}
        \label{intro} 
The understanding of optically thick accretion disks has historically been strongly influenced by analytic models. Thus, one tends to think of thin disks, having in mind the geometrically thin $\alpha$-disk model of \cite{Shakura1973}, with the thickness to radius ratio $H/r \ll 1$; or of thick disks, having in mind the geometrically thick ``Polish doughnuts'' with $H/r > 1$ in the innermost parts, and a funnel through which most of the radiation is emerging  \citep[][]{Jarosz1980}; and of slim disks \citep{AbramSlim} of thickness comparable to the circumferential radius, but usually smaller, $H < r$. In all cases the thickness $H$ can be taken to be the  vertical ($z\equiv r\cos \theta $) coordinate of the photosphere above the equatorial plane of symmetry, and for these three canonical models it happens to coincide also with the characteristic scale-height of density, $h_{\rho}$. A useful discriminant was thought to be given by the luminosity, $L$, of the disk in terms of the Eddington ratio $\lambda \equiv L/L_{\rm Edd}$, with $\lambda < 0.3$ thought appropriate for the thin disks, 
$0.3 < \lambda \lesssim 3$ for the slim disks, and $\lambda  \gg 1$ for Polish doughnuts.

Spectral and flux observations of various black hole sources seem to be in agreement with the predictions of the three models in the stationary approximation. In particular, the multicolor spectrum expected from thin disks yielded good fits to observations of numerous  accreting X-ray binaries \citep{Mitsuda1984}, and the \textsc{xspec} models of thin disk spectra, \textsc{kerrbb} and \textsc{bhspec} \citep{Li2005,Davis2006}, have been successfully used to fit numerous black hole disk spectra at low luminosities. 
Slim disks, physically differing from thin disks in allowing radial advection of the heat generated in viscous dissipation, predict somewhat different spectra (resulting from the temperature being proportional to the radius to the $-1/2$ power) and these were reported to fit both black hole X-ray binary sources of moderate peak luminosity, $\lambda \sim 0.6$ \citep[][]{Kubota2004,Straub2011} and ultraluminous X-ray sources  \citep{Vierdayanti2006,slimulx}. For an up-to-date review of modeling astrophysical sources by slim disks see \cite{Czerny2019}.

While Polish doughnuts were amenable to numerical simulations early on \citep{Igumenshchev2000}, which supported the general picture of a bulbous disk terminating in a cusp, with a fairly narrow axial funnel trough which radiation could have escaped (had it been present in the simulations), the early numerical simulations had nothing to say about the structure of accretion disks at moderate and low accretion rates. Indeed, until recently, with their limited resolution and lack of radiative transport, numerical simulations of sub-Eddington optically thick accretion disks were simply assumed to correspond to the thin or slim disk models. Attention tended to focus on the mechanism of turbulent dissipation and angular momentum transport, as well as the generation and evolution of the internal magnetic field \citep{Brandenburg1995,Machida2003}.

Now that general relativistic magnetohydrodynamic (GRMHD) accretion-disk simulations have come of age and allow global simulations of geometrically thin disks, with \citep[][]{Ohsuga,Sadowski2016a,Mishra2016,Fragile,Jiang} and without \citep{ZhuStone2018,Liska} radiation, it is time to revisit the theoretical properties of sub-Eddington accretion disks. We have simulated disks that are thermally stable in the regime where radiation pressure dominates over gas pressure, and found that at moderate accretion rates they correspond to none of the three analytical models. Instead, in differing respects, they contain elements of each of the three classes of models. The density scale-height of our simulated disks corresponds to those of thin disk models, $h_{\rho}/r \ll 1$. However, the disks are advective, with the photon flux directed mostly inwards close to the inner edge of the disk, more in keeping with the expectations of slim disk models. The overall appearance is reminiscent of thick disks, with the radiation density highest in an axial funnel, with the height of the photosphere satisfying $H/r \sim 1$.

In this letter we discuss a Schwarzschild metric simulation for a ten solar mass black hole ($M=10\,\mathrm{M_\odot}$) with an accretion rate of $\dot{M} = 0.6\,\dot M_{\rm Edd}$, where we define $\dot M_{\rm Edd}= L_{\rm Edd}/(0.057c^2)$ with the usual Schwarzschild efficiency factor $0.057$, and Eddington luminosity $L_{\rm Edd}$.  We are using the gravitational radius, $\mathrm{M}\equiv GM/c^2$, as the unit of length.

\section{Puffy disks}
\label{puffy}

Simulating radiative thin disks, i.e., ones with a low accretion rate, is numerically challenging in magnetohydrodynamics even when the disk is intrinsically stable \citep{Sadowski2016a}, because of the resolution requirements of the magnetorotational instability \citep[MRI,][]{BalbusHawley1998}. Additionally, radiation pressure dominated disks are known to be thermally unstable, with the \cite{Shakura1976} linear stability analysis prediction confirmed for $\alpha$-viscosity disks in fully non-linear, global, radiative, general relativistic, hydrodynamic simulations \citep{Fragile}. This instability has been shown to also hold when the turbulent dissipation is provided by MRI, both in shearing box simulations \citep{Jiang2013}, and in global simulations \citep{Mishra2016}.
\begin{figure*}[t]
\includegraphics[width=1\textwidth,trim=2cm 2cm 0 0]{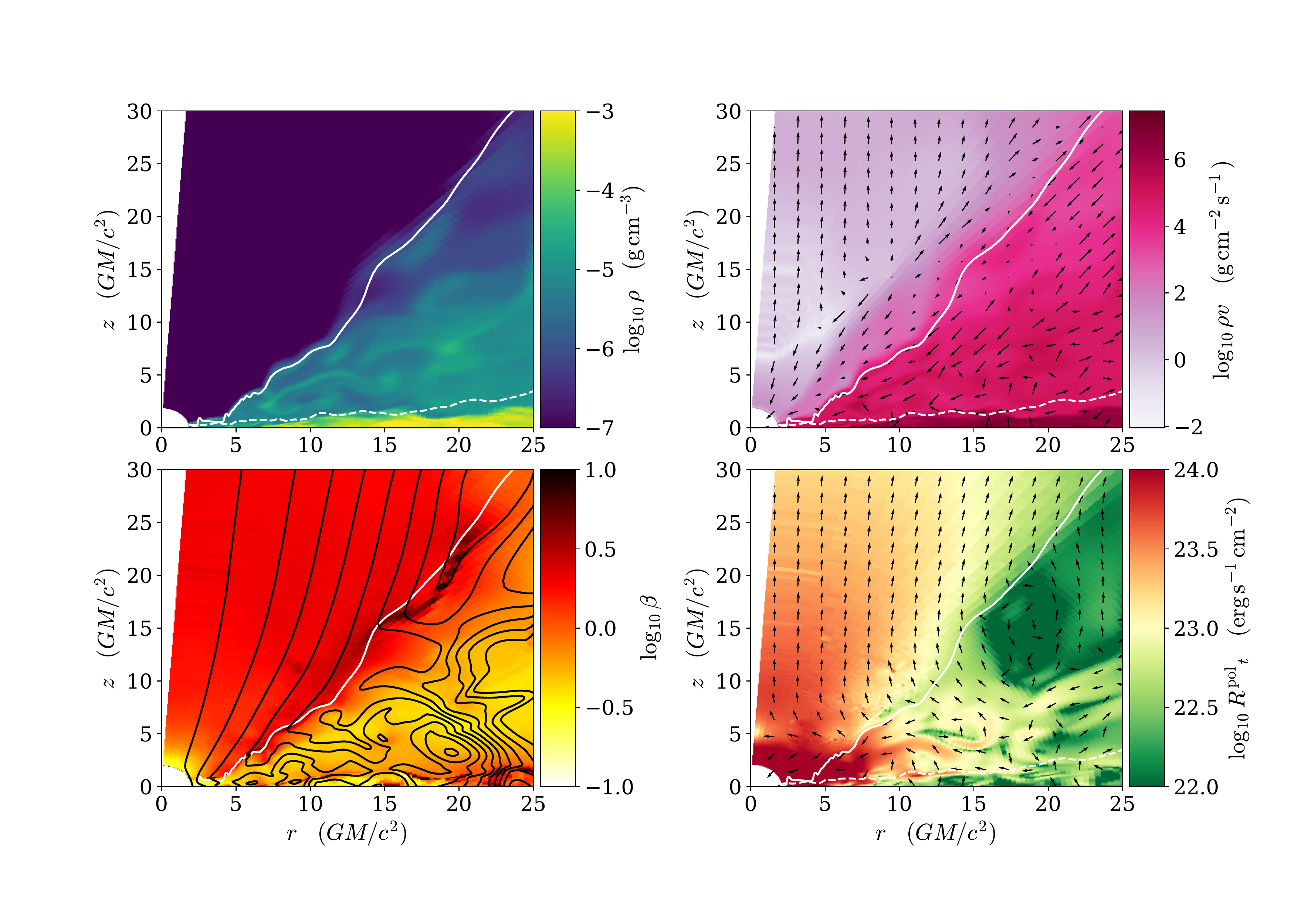}
		\caption{Snapshots showing various physical properties of the puffy disk with $\dot M = 0.6 \dot M_{\rm Edd}$. The location of the photosphere is shown by the solid white line (the optical depth was computed along lines parallel to the $z$ axis), the dashed white line shows the density scale-height, $h_\rho$. {\sl Upper left:} gas density $\rho$.  {\sl Upper right:} gas momentum density $\rho v$ and {vectors of gas velocity in poloidal plane}. 
		{\sl Lower left:} {plasma parameter} $\beta = (p_{\mathrm{rad}} + p_{\mathrm{gas}})/p_{\mathrm{mag}}$ (colours) with contours of the azimuthal component of the magnetic vector potential.  {\sl Lower right:} poloidal component of radiation flux $R^\mathrm{pol}{}_t$ (color) and its direction (unit vectors).
		}
\label{fig:1_snap}        
\end{figure*}

\begin{figure}[ht]
\includegraphics[width=1.0\columnwidth]{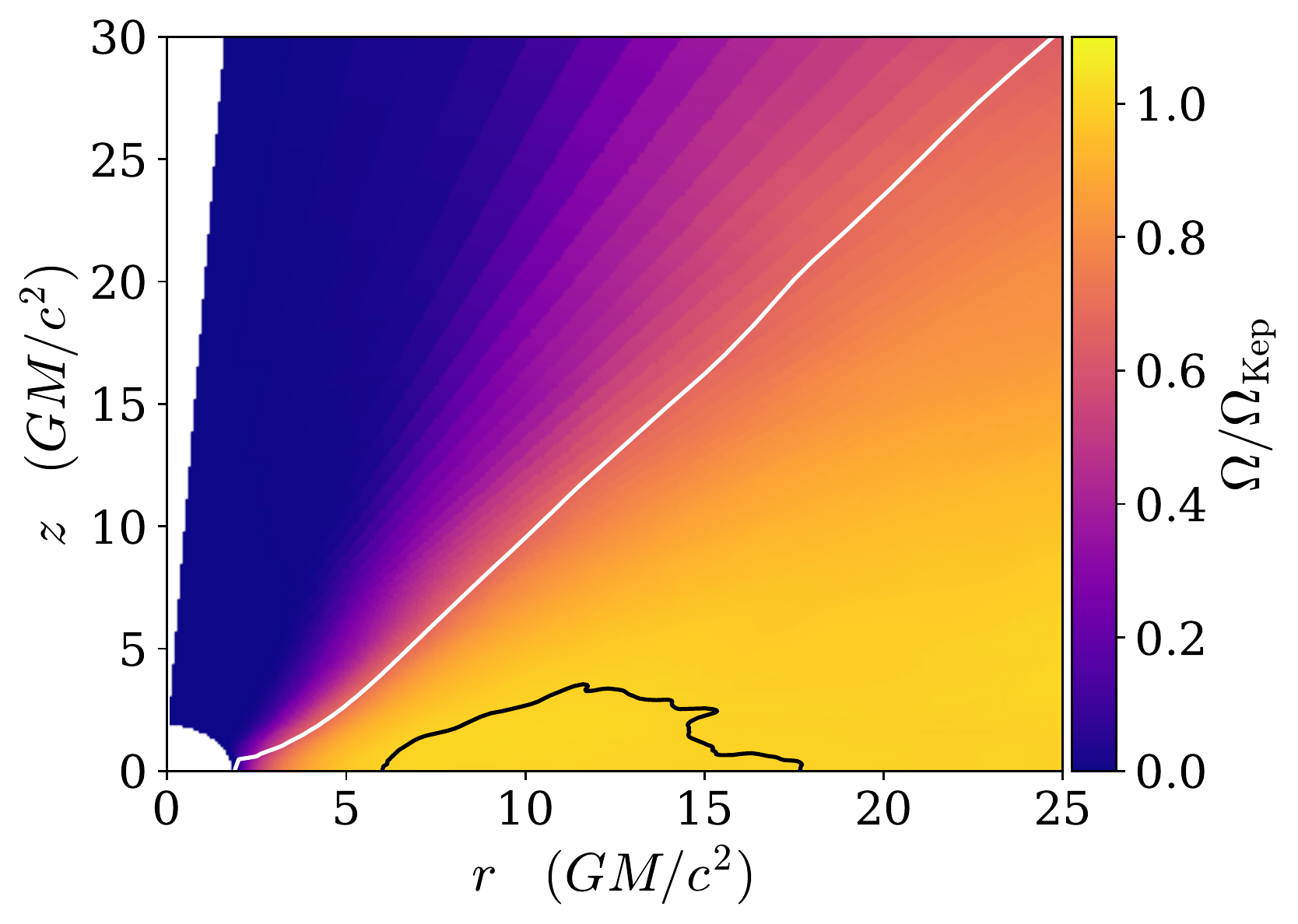}
\caption{Angular velocity frequency $\Omega$ distribution scaled by the  Keplerian value $\Omega_{\mathrm{Kep}} = (GM)^{1/2} R^{-3/2}$ where $R$ is the cylindrical radius. Note that the fluid in the disk is Keplerian essentially all the way up to the photosphere. The black contour encloses the super-Keplerian region.}
\label{fig:3_omega}
\end{figure}

\begin{figure*}[t]
		\includegraphics[width=1\textwidth]{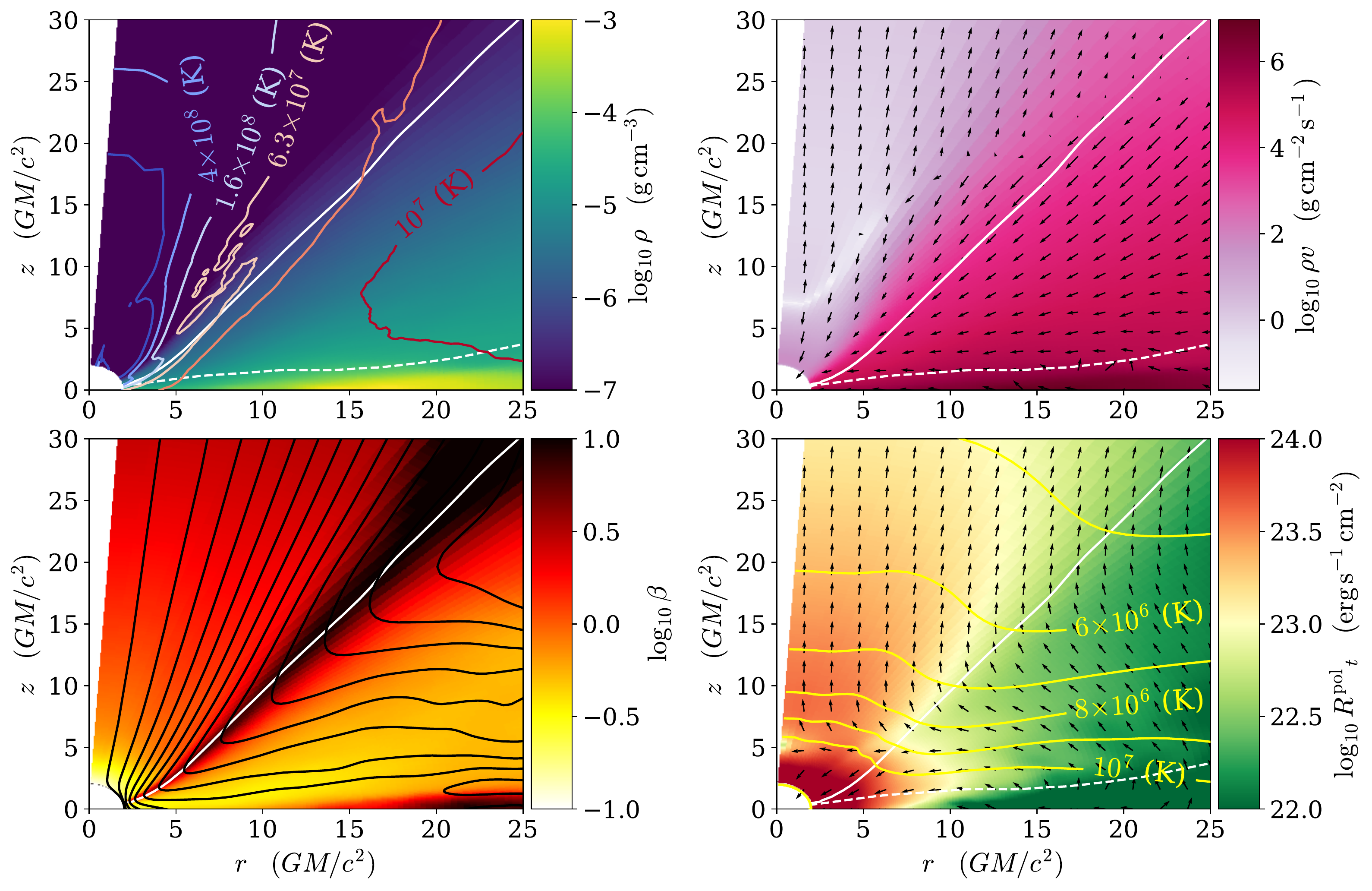}
		\caption{Time-averaged properties of the puffy disk with $\dot M = 0.6 \dot M_{\rm Edd}$. 
		The panels correspond to the panels in Fig.~\ref{fig:1_snap}, with additional information as follows.  {\it Upper left:} contours of gas temperature, from left to right {$T_\mathrm{gas}= 1.0 \times 10^9, 4 \times 10^8, 1.6 \times 10^8, 6.3 \times 10^7, 2.5 \times 10^7, 1.0 \times 10^7$ K}. {\it Lower right:}  contours of radiation temperature, from top to bottom $T_{\mathrm{rad}}=5.0, 6.0, 7.0, 8.0, 9.0, 10.0$ MK.}
        \label{fig:2_avg}
\end{figure*}

It should not be a surprise, then, that radiative GRMHD simulations of accretion flows initialized with a radiation pressure dominated thin disk configuration ended in a rapid collapse of the disk to a state that was too thin for the MRI to be resolved \citep{Mishra2016}. Interestingly, even though most of the gas collapsed to a very thin disk, an optically thick atmosphere permeated by the radiation persisted like the grin of a Cheshire cat, with the photosphere remaining in its original position at a height of nearly $2\,\mathrm{M} \gg h_{\rho}$. In other words, the degeneracy between the density scale-height, $h_{\rho}$, and photospheric half-thickness, $H$, of the disk was broken---they were no longer equal. However, as the simulation had to be terminated this could have been a transient phenomenon and remained but a tantalizing hint of what a ``thin'' disk may really look like. To date, there are no radiative GRMHD simulations of black hole disks at $\sim 0.1$ to $ 0.3\dot M_{\rm Edd}$, so in fact we simply do not yet know what an MHD turbulent black hole disk should look like in theory, in the putative thin disk regime. Following our recent simulations, we are now in a position to describe the structure and appearence of stable $\sim 0.6 \, \dot M_{\rm Edd}$ disks.

We have performed global, 3D, radiative GRMHD simulations of black hole disks at decreasing sub-Eddington accretion rates. To ensure thermal stability \citep{Zheng2011} the disk was made to advect poloidal magnetic field with a significant radial component \citep{Sadowski2016a} from the toroidal mass reservoir often included in disk simulations. Upon evolution of this field through the MRI a major component of the pressure is due to the magnetic field, with the plasma parameter $ \beta = (p_{\mathrm{gas}}+p_{\mathrm{ rad}})/p_{\mathrm{mag}} \sim 1$.
Of the remaining components of pressure,
radiation pressure (hugely) dominates over the gas pressure, and the opacity is dominated by electron scattering, in agreement with \cite{Shakura1973}. The radiation field is evolved in the approximation of the M1 closure scheme \citep{Levermore1984,Sadowski2013}. The simulations were performed with the \textsc{koral} code, and closely follow those of \cite{Sadowski2016a}, which were performed for somewhat higher accretion rates. A detailed description of our simulations can be found in Lan{\v c}ov\'a et al. (in preparation).
Here, we describe the striking features of the simulated stable, sub-Eddington, radiative accretion disks, illustrated by the $0.6 \dot M_{\rm Edd}$ simulation results.

Fig.~\ref{fig:1_snap} shows a meridional-plane snapshot of the density field (top-left panel). It is evident that the accreting gas is concentrated towards the equator, the density scale-height 
\citep[defined as in][]{Sadowski2016a} being $h_{\rho} \sim 0.1 \, r$ ({white dashed} curve). In this sense the disk is geometrically thin. However the potentially observable surface of the disk, i.e., the photosphere (solid white line in all the figures) is at a much greater height, $H \approx r$, just like in a geometrically thick disk. This is not an academic distinction---clearly the shape and position of the photosphere will have significant consequences for the dependence of the flux observed at infinity as a function of the inclination angle to the observer (cf. Section~\ref{discuss}), and for the inferred (so called isotropic) luminosity of the source. Note that very little radiation is released at $r > 15\,\mathrm{M}$, instead, much of it is advected to lower radii and then released into the funnel (bottom-right panel of Fig.~\ref{fig:1_snap}).

 The bottom left panel of Fig.~\ref{fig:1_snap} shows there is a great deal of turbulence in the disk, all the way up to the photosphere.
 Fig.~\ref{fig:3_omega} shows that the disk is Keplerian, also (nearly) all the way up to the photosphere. This would not be surprising for the moderate mass accretion rate of $0.6\,\dot M_{\rm Edd}$, except that our disk is not geometrically thin, and the photosphere is at a height of $z= H\approx r$, so the region of Keplerian rotation is quite thick. Presumably the co-rotation is enforced by the turbulence.

 As the above discussion makes clear, the cartoon separation of the matter in the vicinity of a black hole into a rotating and accreting disk and a quasi-static corona simply does not occur. Instead, the high density core of the disk is sandwiched by a fairly thick ``puffy'' layer of lower-density, magnetized, optically thick plasma participating in the overall flow, and a comoving atmosphere above the photosphere.
\begin{figure}[ht]
	\begin{center}
		\includegraphics[width=1.0\columnwidth]{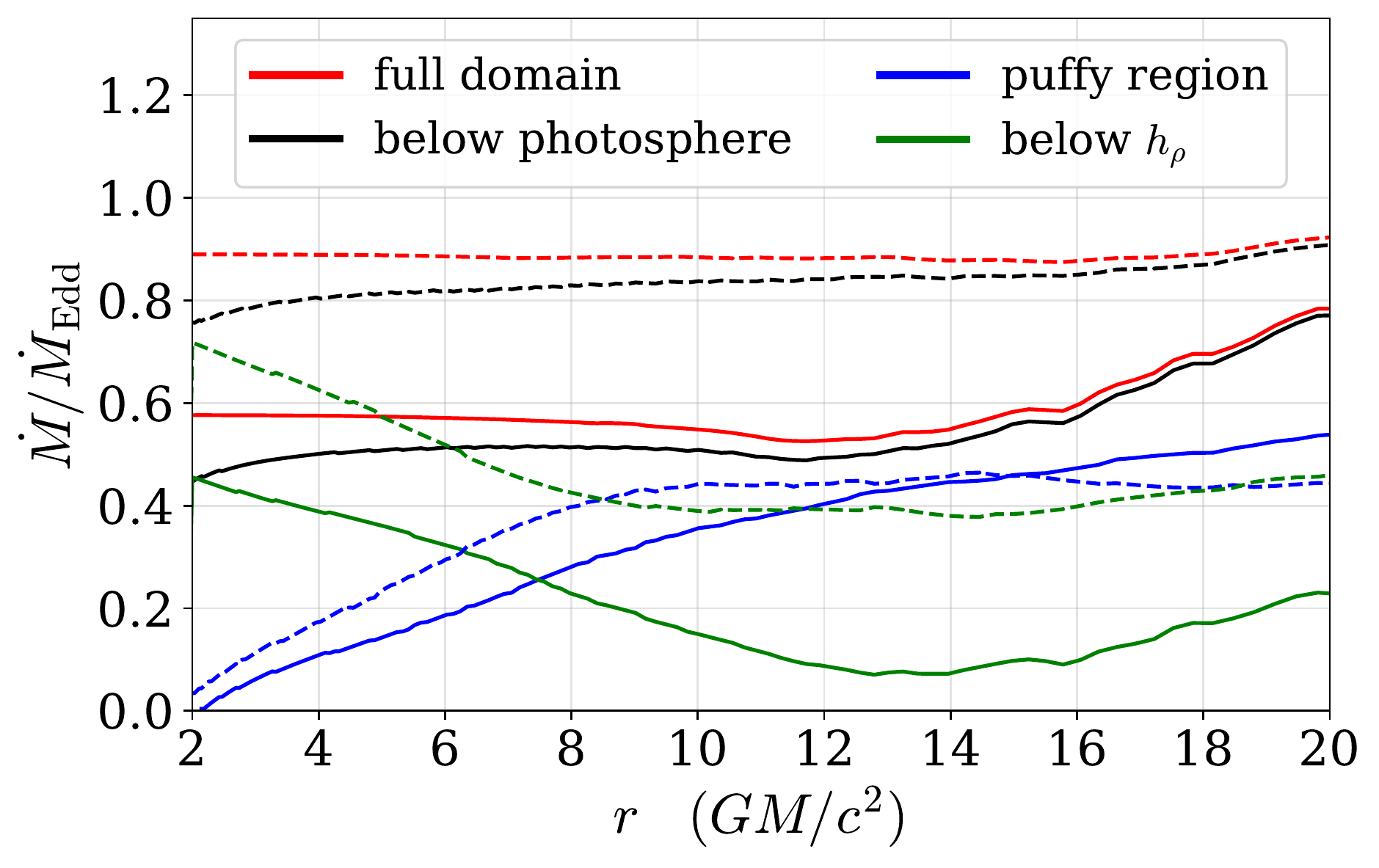}
		\caption{Mass accretion rate as a function of the radius. {\sl Dashed lines:} \cite{Sadowski2016a}. {\sl Solid lines:} the current work. From top to bottom (at larger radii), {\it red:} integrated over the whole range of the polar angle; {\it black:} integrated under the photosphere; {\it blue:} integrated between the photosphere and the density scale-height $h_\rho$; {\it green:} integrated from the equatorial plane to the density scale-height $h_\rho$. At $r \gtrsim 8 \,\mathrm{M}$ most of the accretion occurs in the puffy region between the photosphere and the density scale-height surface $h_\rho$.
		}
	\end{center}
  \label{fig:4_mdot}
\end{figure}

 It is interesting to note that close to the equator there is a region of super-Keplerian rotation (below the black curve in Fig.~\ref{fig:3_omega}) extending from $r \approx 18\,\mathrm{M}$ down to the ISCO (innermost stable circular orbit), at $r=6\,\mathrm{M}$ for a non-spinning black hole. This is qualitatively similar to the expectations in a thick disk, which is sub-Keplerian at distances from the black hole larger than that of the center of the disk (pressure maximum), and super-Keplerian at smaller radii \citep[][]{Jarosz1980}. By contrast, thin disks  are slightly sub-Keplerian except close to the inner edge.\footnote{Specifically, in general relativity thin disks become super-Keplerian close to the ISCO \citep{Jarosz1980}. In Newtonian gravity $\alpha$-disks  are sub-Keplerian down to 1.4 $r_+$, where $r_+$ is the radius where the torque  vanishes \citep[][]{KK}.}

The time averaged quantities are shown in Fig.~\ref{fig:2_avg}.
Note that the motion of the accreting fluid is not confined to the dense core within the scale-height $h_{\rho}$. Quite to the contrary, most of the inflow occurs above the core, and actually dominates it (Fig.~\ref{fig:4_mdot}). The top-right panel of Fig.~\ref{fig:2_avg} shows a great deal of momentum inflow in the vicinity of the photosphere. For $r<10\mathrm{M}$ the fluid is moving towards the black hole in a region extending up to the stagnation surface (clearly visible as a whitish elongated area separating inwards and outwards pointing arrows), which is at least twice as high above the equatorial plane as the photosphere. Thus, there is no evidence for a \cite{Blandford1982} launching of gas, even though the magnetic field lines are inclined outwards (bottom-left panel of  Fig.~\ref{fig:2_avg}) and the fluid is undergoing nearly Keplerian rotation (Fig.~\ref{fig:3_omega}). In the disk the magnetic field is dominated by its azimuthal component, while in the funnel it is dominated by its radial component. Thus, the field changes its character near the photosphere, which results in a minimum of magnetic pressure ($p_{\rm mag}$) there.

It is also apparent (bottom-right panel of Fig.~\ref{fig:2_avg}) that, contrary to what one assumes in the thin and slim disk models (which are  one-dimensional, the equations being height-integrated), radiation that is about to leave the inner disk does not move vertically when still below the photosphere (solid white line), except at the largest radii. The radiation fluid inside the disk flows mostly radially inwards already at $r \approx 10\,\mathrm{M}$,  in contrast to the thin disk model in which the radiation flux vector rotates from the vertical to the radial direction only very close to the ISCO \citep{Bozena}. In the language of slim disks, one would say that in the region defined by $z < 5\,\mathrm{M}$ and $r < 10\,\mathrm{M}$ most of the radiation is advected. Once radiation emerges through the photosphere at $r>5\,\mathrm{M}$, it escapes upward and slightly outwards through an optically thin funnel. There is an accompanying outflow of low-density plasma, apparently pushed out by the radiation.  The radiation released in the inner disk at $z \leqslant 5\,\mathrm{M}$ is lost in the black hole. As predicted by slim disk models, there is a great deal of advection of radiation within the disk. However, in the simulation the advection of radiation occurs also outside the disk,  with inflowing and outflowing radiation regions separated by a ``photon stagnation surface,'' extending from the axis to the photosphere at $z \approx 5\,\mathrm{M}$. For this reason the luminosity of the inner region\footnote{Integrated between the $z$ axis and the photosphere over a spherical surface segment of radius $r =25\,\mathrm{M}$.} is $ L=0.36\,L_{\rm Edd}$, somewhat less than what one would have expected for a thin disk. However, most of this is emitted into a cone subtending much less than the full solid angle (Fig.~\ref{fig:image}), so if observed close to the axis the inferred isotropic luminosity of the source would be significantly larger. 
\section{Discussion}
\label{discuss}
In our radiative GRMHD simulations we have found a new type of black-hole accretion disk which may solve some long-standing problems of accretion disk theory. First, for accretion rates relevant to the bright black hole sources the thin disk solution is known to be radiation pressure dominated in the innermost region, and it has been shown early on that such $\alpha$-disks are disrupted by a viscous instability \citep{Lightman1974}, and are subject to a violent thermal instability \citep{Shakura1976,Pringle1976,Piran1978}; both these instabilities are also present in MRI simulations of radiation pressure dominated  thin disks \citep{Mishra2016}. 
Our solutions are stabilized by magnetic pressure \citep{Sadowski2016a}, in line with the considerations of \cite{Zheng2011}.

Both the thin disk and the slim disk models are based on height-integrated, azimuthally symmetric equations and so are intrinsically one dimensional (1.5D). While slim disks are not subject to the thermal instability of thin disks, this property is not confirmed by simulations. In the absence of stabilizing magnetic field all global 3D radiative GR simulations of sub-Eddington disks in the radiation-pressure dominated regime lead to disk collapse \citep{Sadowski2016a,Mishra2016,Fragile}. 
We surmise that the presence of advection alone is not enough to stabilize sub-Eddington disks. Furthermore, while capturing a great deal of the relevant physics, the thin and slim disk models, being intrinsically one dimensional, do not allow for realistic meridional flow patterns, such as the equatorial backflows found  for thin disks analytically \citep{Urpin,KK}, numerically \citep{Regev,Philipov2017}, and in hydrodynamic simulations of thick disks \citep{Igumenshchev2000,Lee2002}. For the same reason they do not allow for non-trivial patterns in the flow of radiation, such as those observed in our 3D simulations (bottom-right panels of Figs.~\ref{fig:1_snap},\,\ref{fig:2_avg}).

\begin{figure}[ht]
		\includegraphics[width=1.0\columnwidth,trim=0.4cm 0 0.1cm 0]{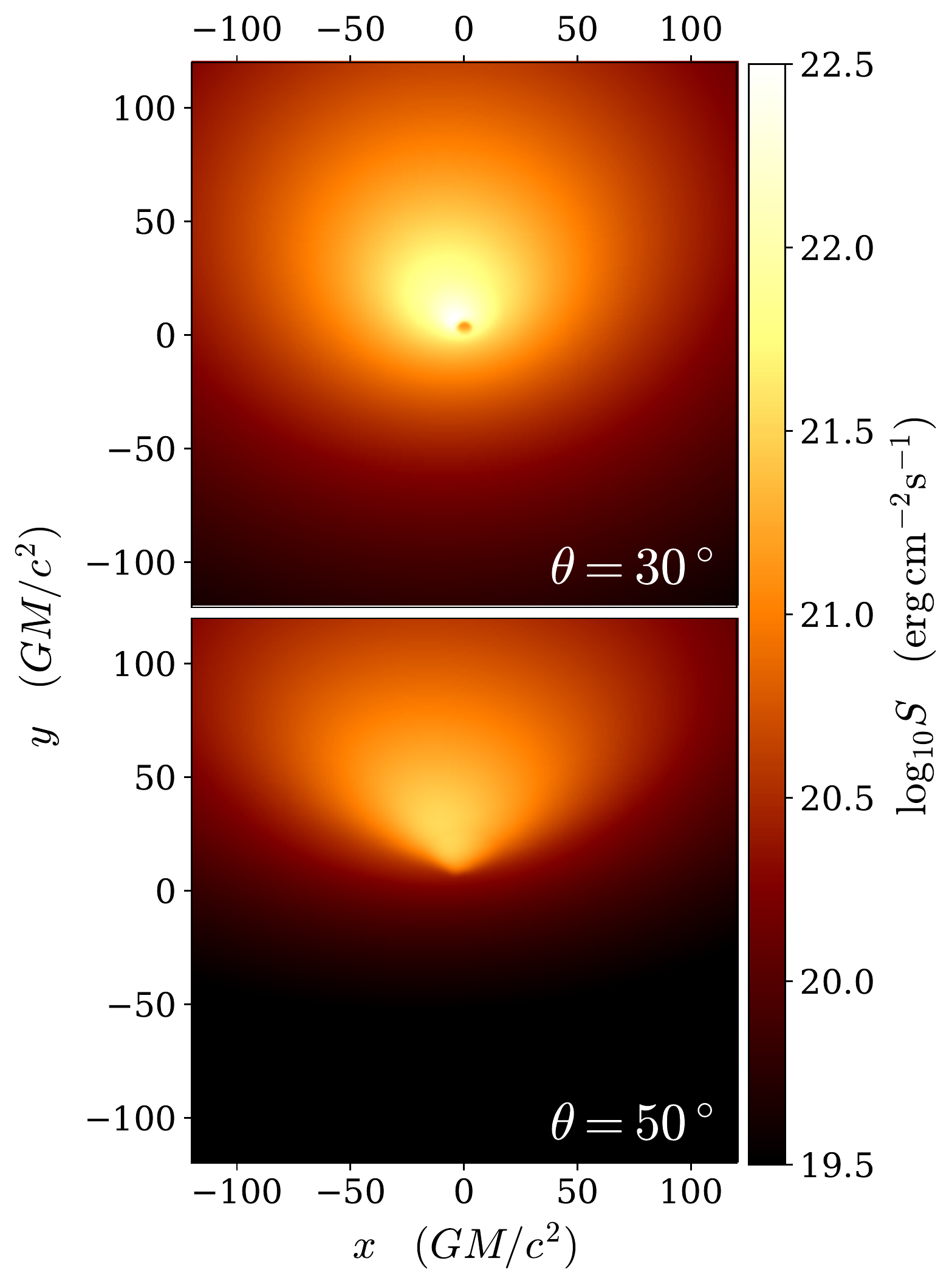}
		\caption{The ray-traced image of the inner part of the puffy disk for line of sight inclination $\theta$ of $30^\circ$ and $50^\circ$ to the axis.}
\label{fig:image}
\end{figure}
From the observational side, studies of AGNs have recognized the need for ``elevated'' disks, thicker than the standard thin disk solutions \citep{Begelman2017}. Our models respond to this need, and correspond to the recent magnetized disk models of \cite{Mishra2019}. The latter work, a Newtonian simulation without radiation, has superior resolution and shows a spiral pattern of accretion, which has not been obtained in our limited azimuthal domain (c.f, the Appendix). Our findings of accretion predominantly occurring above the high density core agree with those of \cite{ZhuStone2018} and \cite{Mishra2019}. In fact, for the first time in an MRI simulation, \cite{Mishra2019} find an equatorial backflow, their meridional accretion flow pattern being reminiscent of the analytic $\alpha$-disk results of \cite{KK}. We do not find clear evidence of such backflow in our simulation, it would be interesting to check whether its absence is a result of the strong radiation field or strong gravity.

We also expect puffy disk spectra to be different from those of the slim disks (for instance, small density and high temperature imply a large color correction to the thermal spectrum) and to this end we will be post-processing our simulation to compute the spectra at infinity (Lan\v{c}ov\'{a} et al., in preparation). 
The computed appearance at infinity of the puffy disk  (Fig.~\ref{fig:image}) differs from the usual image of a warped thin disk with a hole in it. At small and moderate inclinations (up to about $40^\circ$ away from the axis) a wide bright funnel is clearly apparent with a darker spot just above the black hole (which itself is partially obscured by the hot gas). At larger inclinations, the near side and the bottom of the funnel are shadowed by the optically thick disk.

The $0.6\, \dot M_{\rm Edd}$ accretion rate in the reported simulation may correspond, close to the peak of their brightness, to such black hole sources as LMC X-3, for which an \textsc{xspec}  model of slim disk spectra was developed \citep[\textsc{slimbb},][]{Straub2011}, a model which may now need to be supplanted.\footnote{Several other slim-disk spectral fitting routines are available, e.g., \textsc{slimulx} \citep{slimulx}, and others at  https://heasarc.nasa.gov/xanadu/xspec/models/slimdisk.html
\citep[based on][]{Kawaguchi2003}, and in the GitLab project https://projects.asu.cas.cz/bursa
/slimbh/tree/master.}

The puffy disk is  expected to have different timing properties than the thin disk and its oscillation eigenmodes may differ from those of diskoseismology \citep{Wagoner1999,Kato2001} and of accretion tori \citep{Rezzolla2003,Blaes2006,Mazur2016,Mishra2017}.  This may be relevant to the theory of black hole QPOs \citep{Remillard2006}. The phenomenology of state transitions in black hole X-ray binaries may also need reinterpretation.

\acknowledgments

	~~~~~~~~~~~~~~~~~~~~~~~~~~~~~~~
	
The authors thank Andrew Chael and Brandon Curd for support with the \textsc{koral} code.	The computations in this work were supported by the \textsc{PLGrid Infrastructure} through which access to the Prometheus supercomputer, located at ACK Cyfronet AGH in Krak\'{o}w, was provided. This work was supported in part by the Polish NCN grant 2013/08/A/ST9/00795, the \textsc{inter-excellence} project No. LTI17018 aimed to strengthen international collaboration of Czech scientific institutions, and the Black Hole Initiative at Harvard University, which is funded by grants the John Templeton Foundation and the Gordon and Betty Moore Foundation to Harvard University.
M.A. acknowledges the Polish NCN grant 2015/19/B/ST9/01099, M.A., D.L and G.T. the Czech Science Foundation grant No. 17-16287S, and D.L. the student grants SGS/13/2019 and MSK/03788/2017/RRC.


\appendix
Here,  we provide some  details of the numerical simulations.
The essential technical description of the \textsc{koral} code  is given in \cite{Sadowski2013,Sadowski2014}.
We extended the $\sim 0.8 \dot M_{\rm Edd}$ results published in \cite{Sadowski2016a} to obtain a family of stable solutions at lower accretion rates. In our current work we only use reservoirs of mass with quadrupole topology of the magnetic field, resulting in strongly magnetized stable solutions. Initially, the toroidal reservoir extends from $r=42\,\mathrm{M}$ to $r=500\,\mathrm{M}$. For initial conditions of disks with successively lower accretion rates we are using the results of a previous simulation at larger $\dot M$,  scaling down the gas density, internal energy, radiation energy density,  and magnetic field by a constant factor. This preserves temperature, the gas to magnetic pressure ratio, and $\beta$. 

The $0.6 \dot M_{\rm Edd}$ simulation has achieved inflow equilibrium out to $r \approx 20\,\mathrm{M}$ in a run of duration   $t= 15000\, GM/c^3$. 
Throughout the domain we are using horizon-penetrating,  modified Kerr-Schild  coordinates. The simulations demands an enormous amount of computational time even with a very effectively configured grid, strongly concentrated around the equatorial plane, so we use only a $\pi/2$ wedge with periodic boundary conditions in the  azimuthal direction. It has been shown by \cite{Sadowski2016a} that in the disk properties that we discuss in this {\sl Letter} there is no significant difference between 3D simulations over a $\pi/2$ wedge and a domain spanning the full $2\pi$ azimuthal angle. The resolution of the grid ($N_r \times N_{\theta} \times N_{\phi}$) is  $320 \times 320 \times 32$, which is sufficient to resolve the MRI \citep{BalbusHawley1998} with quality factors $\langle Q_\theta \rangle \approx 25$ and $\langle Q_\phi \rangle \approx 20$ in the accretion flow.

The inner disk ($r<25\,\mathrm{M}$) simulation output (temperature, density, velocity, and magnetic field), and its extrapolation for $25\,\mathrm{M} \leq r < 500\,\mathrm{M} $, was postprocessed to generate the ray-traced images of the disk (Fig.~\ref{fig:image}).  To this end, the radiation field was solved for with the \textsc{heroic} code \citep{Heroic,Narayan2016}, including bremsstrahlung, synchrotron, and Comptonization processes.

\vfill\eject
\end{document}